\documentclass[12pt,preprint]{aastex}
\usepackage{epsfig}

\usepackage{emulateapj5}

\def\be{\begin{equation}}
\def\ee{\end{equation}}
\def\bc{\begin{center}}
\def\ec{\end{center}}
\def\beq{\begin{eqnarray}}
\def\eeq{\end{eqnarray}}

\def\nbl{N_{\rm BL}}
\def\Nu{N_{\rm U}}
\def\ne{n_{\rm e}}
\def\Nobj{N_{\rm obj}}
\def\Cr{C_{\rm r}}
\def\Co{C_{\rm o}}
\def\mag{^{\rm m}}

\shorttitle{Search for the real sample}
\shortauthors{Stern \& Poutanen}

\slugcomment{Accepted to ApJ Letters}

\begin{document}

\title{Blind search for the real sample: \\
Application to the origin of ultra-high energy cosmic rays}

%\title{``Search for the real sample'': A new method for a blind statistical test. \\
%Application to the origin of ultra-high energy cosmic rays}

 \author{Boris~E.~Stern\altaffilmark{1,2} and Juri~Poutanen\altaffilmark{3} }

\affil{Astronomy Division, P.O.Box 3000, 90014 University of Oulu, Finland; \\
 stern@bes.asc.rssi.ru, juri.poutanen@oulu.fi}

 \altaffiltext{1}{Institute for Nuclear Research, Russian Academy of Sciences,
7a, Prospect 60-letija Oktjabrja, Moscow 117312, Russia}
\altaffiltext{2}{Astro Space Center of Lebedev Physical Institute,
Profsoyuznaya 84/32, Moscow 117997, Russia}
\altaffiltext{3}{Corresponding Fellow, NORDITA, Copenhagen}

\begin{abstract}
We suggest a method for statistical tests which does not suffer from
{\it a posteriori} manipulations with tested samples (e.g.
cuts optimization) and does not require a somewhat obscure procedure of
the penalty estimate. The idea of the method is to hide the real
sample (before it has been studied) among a large number of artificial
samples, drawn from a random distribution expressing the null
hypothesis, and then to search for it as the one demonstrating the strongest hypothesized effect.
The statistical significance of the effect in this approach is
 the inverse of the maximal number of random samples at which the search was successful.
We have applied the method to revisit the problem of correlation between the
arrival directions of ultra-high energy cosmic rays  and BL Lac objects.
No significant correlation was found.
\end{abstract}


\section{Introduction}

 Communications about effects detected at a marginally significant level
constitute a considerable fraction of all scientific results.
The scientific society usually treats such communication with
skepticism. Indeed, too many marginally significant effects
have not withstood the data accumulation.

 High energy astrophysics gives a number of instructive examples of searches for
marginally significant effects. Indeed, there are many detection of particles or
transient gamma-ray events whose sources (i.e. objects known from observations at other
wavelenghts) are unknown. This stimulates intensive searches
of various correlations between different classes of events and objects. For example, there
is a number of works reporting detections of correlation between locations of gamma-ray
bursts (or their sub-samples) and various objects: galaxy clusters
\citep{kp96}, Galactic plane \citep{bel97},
and the local galactic arm \citep{kkt97}.
None of these results has been confirmed.
Another similar area is ultra-high energy cosmic rays (UHECRs) and searches for
their hypothetic sources. A claim of significant autocorrelation in the
arrival directions of UHERCs detected by the Akeno Giant Air Shower Array (AGASA)
\citep{h96,t99} motivated searches of cross-correlations
between UHECRs and various astrophysical objects. Particularly there were
reported statistically significant cross-correlation signals between UHECRs and BL Lac
objects (\citealt*{tt01}, hereafter TT01, but see   \citealt*{efs03}),
super-galactic plane \citep{uch00}, radio-loud compact
quasars \citep{vir02}, highly luminous, bulge-dominated
galaxies (presumably, nearby quasar remnants, \citealt*{t02}) and
Seyfert galaxies \citep{ury04}.

The reason for abundance of detected correlations is quite evident: a number of various possible
effects, which have been searched for with statistical methods, is large
and it is not surprising that some of them
demonstrate a marginally significant signal just by chance. The situation is
even worse because typically a probed effect is somewhat uncertain and
the researcher tries different versions of the hypothesis, varying parameters
and applying various cuts to the data samples.
 This means that the
researcher performs a number of tests
of the same effect which are neither independent nor completely dependent.
These numerous trials, again, increase the probability
to observe a signal in one of the trials by chance and the analysis
of this kind of bias is difficult. We illustrate this problem in Section \ref{sec:inter}
and in Fig.~1.

Does it mean that one should reject the possibility to manipulate the data
samples with cuts and parameters? A blind test
when all cuts and parameters in a statistical test have been set and
motivated {\it a priori}, is a good style.
But there are many situations when such {\it a priori} definition of a test
is very problematic and the investigator sometimes really needs the rights to
vary the testing procedure and to see what will happen.

In principle, the researcher can account for these numerous trials
using random samples, representing the null hypothesis.
%Below we describe
%two approaches for estimates of the significance with randomly generated
%data samples.
%First of these approaches is more or less known, particularly it
%has been used by TT01.
%The second approach is, up to our knowledge, new (at least in astrophysics)
%and its suggestion is the main objective of this work.
Often it is done in the following way (see e.g. TT01).
%In the first approach
The investigator prepares a large array of $N$ random
samples,  $d^i$, and does the same estimate
of the effect for each of these samples  as he does for the real
sample,  $d^0$, in each statistical trial.  Let a statistic associated
with a confidence of the effect  (e.g.   $1-p$, where $p$ is the probability
to obtain the result from null hypothesis) be $S^i_j = F( d^i, C_j )$, where $C_j$ is a set
of cuts and/or analysis parameters from the universe $C$ of all cuts and
parameters. First, one finds the maximum for the real sample
$S^0_{\max} = \max \{F(d^0, C_j)\}$    which is reached at   $j = j_0$.
Then, one performs similar search for random samples
$S^i_{\max} = \max \{F(d^i, C_j)\}$.
The significance can be defined as the fraction of random
samples satisfying the condition $S^i_{\max} > S^0_{\max}$. This value differs
from the straightforward (uncorrected for numerous trials) estimate of the
significance by the ``penalty factor''
$N(S^i_{\max} > S^0_{\max})/N(S^i_0 > S^0_{\max})$, where
$S^i_0 = F(d^i, C_{j_0})$  and $N(\bullet)$  is the
number of samples satisfying a condition $\bullet$.
 This procedure is sufficient if
(i) the investigator follows the above procedure precisely;
(ii) the investigator does not use the {\it a posteriori} information on the
real sample for the planning of the investigation strategy.

We would like to notice that both conditions are not so easy to satisfy
once the investigator studied the real sample and feels which combination of
cuts or model parameters will provide the most significant signal. Then he
can find the most favorable trial   intuitively, avoiding a large number of
unfavorable ones. In other terms, the investigator can   introduce
a bias in the choice of $C_j$
and   overestimate the significance
of the effect using {\it a posteriori} knowledge. We should emphasize that the
investigator can introduce such bias not deliberately.  This is a serious
disadvantage of the approach. Such kind of bias is difficult to trace and
 we consider   this method to be insufficiently credible.

In this work, we suggest a new approach   giving a simple way of avoiding
this ``pressure'' of the {\it a posteriori}
information. The investigator can  hide the real sample inside
a large array of random null hypothesis samples {\it prior to any data
analysis}. Now we have a single array of $N$
samples $d^i$. One of them is real (the investigator does not know which), other are
random. The problem is thus inverted: instead of
confirming the hypothesis using the real sample,
the investigator must find the real sample in the array using the hypothesis
that the verified effect exists, i.e.
find $i_{\max}$ corresponding to the maximum value of $S^i_j = F(d^i,C_j)$. This is a
blind test: the investigator does not know where is the real sample and
he can feel free to perform numerous trials.
If the investigator finds the real sample, the significance of the effect is
just the inverse of the number of samples in the array.

An alternative to our method is the cross-validation method, where
the search for an effect is carried out on {\it a fraction} of
the data sample. Therefore it is  less sensitive.
Below we  demonstrate our method applying it to the problem of  UHECRs -- BL Lacs
correlation.

\section{Procedure}

\subsection{Catalogs}

We used the AGASA sample of UHECRs with 58 events above $4 \times 10^{19}$ eV
and a catalog of \citet{vv03} containing 876 BL Lac objects.
We do not combine the AGASA sample with the data from other experiments because
other samples are
smaller and problems associated with the non-uniform structure of a joint sample
would overweight the statistical gain.
The BL Lac catalog has been cut in declination at $-10\degr$ and was subject to various
brightness cuts. We also tried a sub-catalog of {\em confirmed} BL Lacs
which includes 491 objects. Actually  it is not clear which catalog
is more relevant (TT01 used a confirmed sub-catalog)
and therefore we try both variants.

\subsection{Null hypothesis and random samples}

\label{sec:sample}

 Null hypothesis in our case is just
the isotropic distribution of arrival directions of UHECRs
convolved with the AGASA exposure
function. The latter is a function of declination and does not depend on right ascension.
This provides a simple way to prepare random, null-hypothesis samples avoiding
possible uncertainties in the latitude exposure function: to sample the right ascension
uniformly keeping the actually observed declination for each event.
We, nevertheless, have dispersed the declinations of UHECRs by $\pm 3\degr$
around their real values in order to destroy a possible small-scale latitude
correlation, if the latter exists.
Such small dispersion does not distort a much wider exposure function.

 When performing the test we have distributed roles: one of the coauthors
acts as an ``investigator'', another plays a role of ``examiner''.
Examiner has prepared
an array of 99 random samples as described above and
inserted the real sample into the array
keeping the sequential real sample number in secret from the investigator.
He did not participate in the data analysis until the investigator
made his final choice.

\subsection{Measure for the correlation signal}

 We used usual two-point correlation
function counting the number $n$ of UHECRs within angle $\delta$ from any BL Lac of a given
catalog. Then we compare this number with expectation $\ne$ for the null hypothesis:
\begin{equation} \label{eq:expect}
\ne = \nbl \Nu \frac{1- \cos\delta}{1-\cos(-10{\degr})} ,
\end{equation}
where $\nbl$ is the number
of BL Lacs in the catalog, $\Nu = 58$ is the number of UHECRs, $-10{\degr}$ is the
declination cut on BL Lacs. Note that this expectation implies an isotropic distribution
of at least one sample. This is not the case because the AGASA sample has a latitude
anisotropy
and BL Lac catalog is anisotropic respectively to the galactic plane (selection effect)
and the cosmological large-scale structure.
A more accurate estimate differs from that given by equation (\ref{eq:expect}) by a
factor
\begin{equation} \label{eq:fcorr}
F = {\Sigma_{i=1}^{\nbl} \xi(\theta_i)\over \nbl \langle\xi\rangle },
\end{equation}
where $\xi(\theta)$ is the AGASA exposure function. The exposure function
 depends on particle energy and is hardly known better than one can
 extract from the latitude distribution of detected UHECRs.
\citet{t99}  use a
polynomial fit to the observed latitude distribution of events above $10^{19}$ eV. We
 prefer to use the observed distribution of the available AGASA sample
 (above $4\times 10^{19}$ eV)
in a form of histogram in
$\cos\theta$ with the bin width 0.1 since this is a simplest option that can be
easily reproduced.

 Factor $F$ depends on the BL Lac catalog and
therefore on cuts. According to our estimates with equation (\ref{eq:fcorr}), $F$ is close to 1 for
radio-bright
objects and  $\sim 1.2$ for optically-bright objects (probably due to anisotropy caused by
galactic absorption).
 We introduce the measure of the signal,
$p$ (which depends on $\delta$ and cuts in the BL Lac catalog), as the probability to
sample $n$ or more hits from the Poisson distribution at expectation $F \ne$.

Note that for autocorrelated samples the distribution of $n$ is not Poisson,
therefore this measure is not exact. In order to correct this probability
for the actual autocorrelated distribution of BL Lacs on the sky, we
perform Monte-Carlo simulations using a large number of random UHECR samples.
The maximal disagreement between the Poisson and Monte-Carlo probabilities
is by a factor of 2. Thus, we use the Poisson probability, $p$,
for preliminary estimates and recalculate the probability
for the  leading samples (given in Table~1) with Monte-Carlo simulations.

%%%%%%%%%%%%%%%%%%%%%%%%%%%%%%%%%%%%%%%%%%%%%
\footnotesize
\begin{center}
{\sc TABLE 1\\
Samples of UHECRs demonstrating the most significant correlation with BL Lacs}
\vskip 4pt
\begin{tabular}{cccccc}
\hline
\hline
R or O$^{\rm a}$ & ID$^{\rm b}$ & $\Nobj$ $^{\rm c}$ &
 $\Cr$ or $\Co$ $^{\rm d}$ & $\delta^{\rm e}$ & $p\times 10^4$ $^{\rm f}$\\
      &         &        &  (Jy or $V$)    & (deg)    &      \\
\hline
 \multicolumn{6}{c}{All quasars} \\
  R & 90 & 256 & 0.04   & 2 & 2.6  \\
  R & 40 & 139 & 0.16    & 3 & 3.2  \\
  R & 11 & 35  & 0.79    & 2 & 5    \\
  O & 90 & 153 & 17.5   & 2 & 3.12 \\
\multicolumn{6}{c}{Confirmed BL Lacs} \\
  R & 11 & 6   & 0.79    & 2   & 1.1  \\
  R & 4  & 197 & 0.02    & 1.5 & 3.47 \\
  R & 90 & 6   & 0.79    & 3   & 8    \\
  O & 4  & 118 & 18      & 1.5 & 1.15 \\
\hline
\end{tabular}
%%%%%%%%%%%%%%%%%%%%%%%%%%%%%%%%%%%%%%%%%%%%%
\end{center}
{
$^{\rm a}$  Cut applied in radio, R, or optical, O, brightness \\
$^{\rm b}$  Identification number of a sample giving
the strongest correlation signal  \\
$^{\rm c}$ Number of objects passing the cut \\
$^{\rm d}$ Optimal cut in 6 GHz radio flux or visual magnitude \\
%for a given sample \\
$^{\rm e}$ Optimal correlation angle\\
$^{\rm f}$ Significance level
}
%\centerline{}
\normalsize
\medskip
%%%%%%%%%%%%%%%%%%%%%%%%%%%%%%%%%%%%%%%%%%%%%

\section{Search for the best-correlating sample and its results}

 Optimizing cuts in all existing parameters   we can fit
a BL Lac catalog to any set of locations in the sky so that it will
demonstrate a highly significant correlation (see Sect. 4). Therefore, if our
objective is to find the real sample, we have to try the most relevant
cuts. The apparent radio- or
optical brightness of objects (represented in the catalog by their observed radio
flux density measured in Jy and the visual magnitude $V$) seem to be
good  indicators of particle acceleration to ultra-high energies.
To avoid ``over-optimization'' of random
samples in two-dimensional scan, we performed two separate scans:

\begin{enumerate}
\item We optimized cut $\Cr$ in the  6 GHz  radio flux within the limits
0.01 Jy $< \Cr <$ 2 Jy, varying it with the step 0.1 in decimal logarithm. No
cuts in optical brightness was applied. This scan is marked with letter R in
Table~1.

\item We optimized cut $\Co$ in visual magnitude within the range from $V=12$ to $V=24$
with the step $\Delta V=0.5$. No cuts in radio  flux was applied and we excluded objects
with no data on their radio brightness. This scan is marked with letter O in
Table~1.
\end{enumerate}

The proper correlation angle $\delta$ is  somewhat uncertain.
The most significant correlation should not certainly appear at a correlation angle
equal to $1 \sigma$ experimental error (the latter depends on the particle energy).
If UHECRs are charged, then the correlation could appear at $\delta$ corresponding
to a typical angle of particle deflection. We optimized $\delta$ between $1{\fdg}5$ and
$5{\degr}$ with the step $0{\fdg}5$.
The samples that give the most significant correlation  are listed in Table~1.
In addition,
we also tried a scan over the intrinsic radio luminosity as was done in TT01. The strongest effect gave
sample \#11: $p = 4  \times 10^{-4}$ with 25 intrinsically brightest BL Lac
objects and $\delta = 3\degr$.

With these results at hand, the investigator had to make a choice concerning the
real sample.
All best samples (except \#4) have a reasonable value of optimal
$\delta$ ($2{\degr}$ and $3{\degr}$),
which is close to the angular resolution of AGASA of $2{\fdg}3$.
Finally,  the ``investigator''
used sample \#11 as the first choice. The second option was sample \#90.

The second task is the test for autocorrelation of the UHECR arrival directions.
It was performed with the
same array of random samples before the ``investigator'' was informed about
the results of his choices in the first test. The autocorrelation signal
is estimated in a similar way as described above for the cross-correlation
signal:
\begin{equation}
\ne = \frac{\Nu (\Nu-1)}{2} { 1-\cos\delta \over 1-\cos(-10^{\degr})},\quad
\strut\displaystyle
F = {\Sigma_{i=1}^{\Nu} \ \xi(\theta_i) \over \Nu \langle \xi \rangle } ,
\end{equation}
where factor $F=1.4$.

Now, sample \#67 showed maximum  signal of
$p = 0.5 \times 10^{-3}$ at $\delta = 2{\fdg}5$ (8 hits).
The second  sample showing strong  autocorrelation
was \#30 with $p = 1.7 \times 10^{-3}$
at $\delta= 2{\fdg}1$. The choice of the investigator was \#67.

The real observed sample of UHECRs  had sequential number \#67.
Therefore the test at 99 per cent confidence
level was unsuccessful for UHECRs--BL Lacs correlation and successful
for UHECRs autocorrelation. Then we checked sample \#67 for the cross-correlation
with BL Lacs by varying $\Cr$ and have not found any significant signal.

\section{Interpretation of the results}
\label{sec:inter}

We can confirm that the autocorrelation signal in AGASA sample with
the given energy threshold has a significance of at least $10^{-2}$. To
find the significance level we would have to vary the size of the
random array and to find the limit when we are able to find the real
sample. This objective is beyond the scope of this work. Probably,
according to the correlation signal in the second best sample, the
significance is around $3 \times 10^{-3}$ in agreement with
 \citet{fw04}. One should notice, however, that this
result refers to a specific sample with the energy cut of $4 \times 10^{19}$
eV (see  \citealt*{fw04}, for the discussion).
To estimate the significance of real autocorrelation one has to perform the
same procedure with an untruncated sample of UHECRs varying the energy cut
in a reasonable range.

Our negative result on cross-correlation with BL Lacs
does not mean   that we have found a   quantitative  disagreement with the results of
TT01. They have found a positive signal with an another catalog of the
confirmed BL Lac objects. Their cuts were: $z > 0.1$ or unknown, $\Cr=0.17$ Jy,
$\Co=18\mag$. At these cuts the positive signal still exists at
$p=1.9\times10^{-2}$ and $\delta = 2\fdg5$ (with factor $F = 1.24$,
 see Eq.~\ref{eq:fcorr}) and the real sample  \#67 is
the second significant among 99 random samples (having similar significance
with three other samples  including sample \#11).

\bigskip
%\begin{figure}
\centerline{\epsfxsize=8.4cm \epsfbox{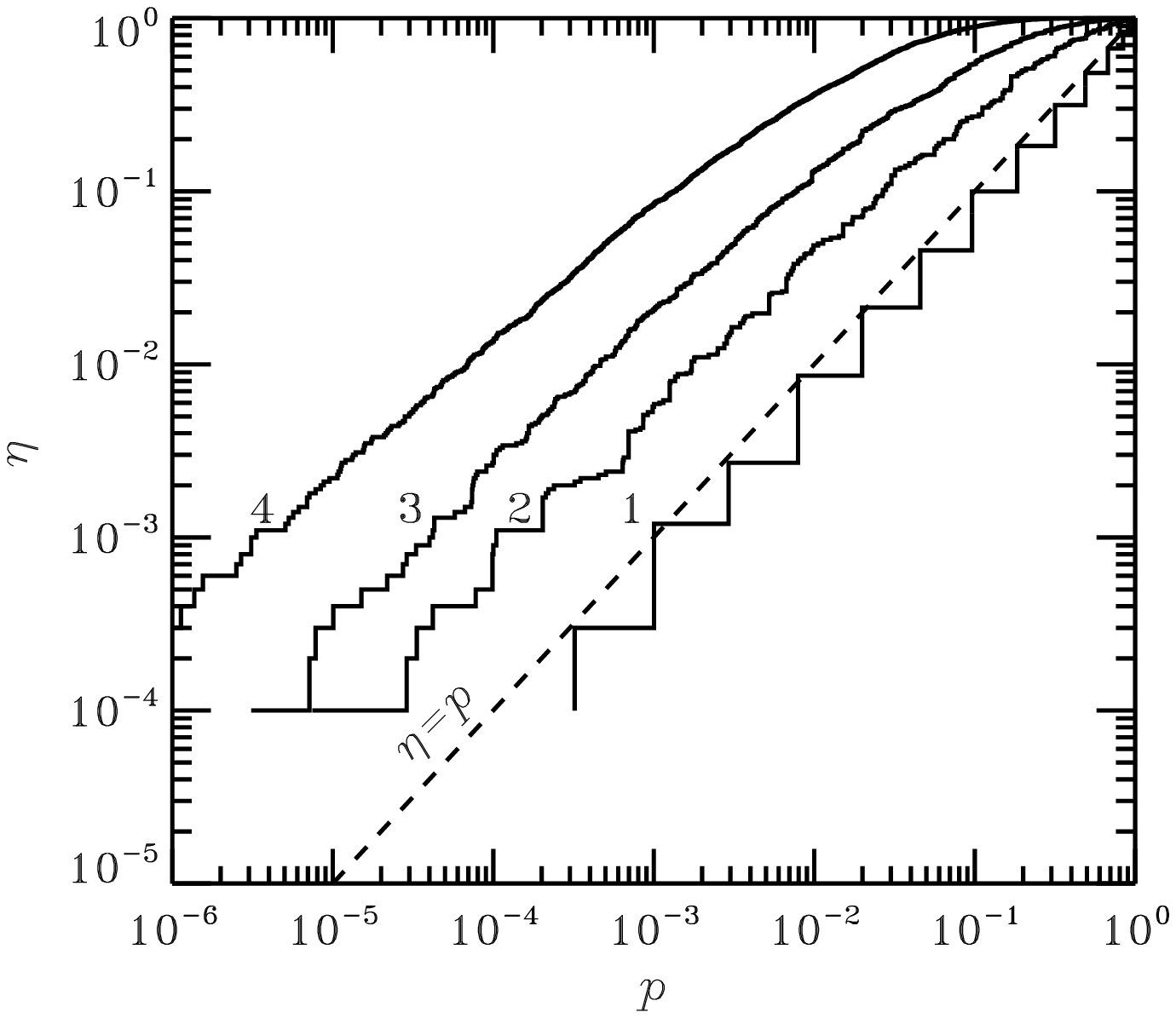}}
\figcaption{
%\caption{
The fraction of $10^4$ simulated random UHECRs samples, $\eta$,
demonstrating a higher significance
level for the ``correlation signal'' with the BL Lac
catalog of \citet{vv03} (876 objects)
than a given value $p$ for different cut optimization. From lower to higher curves:
1 -- no cuts optimization with $\Cr=0.2$ Jy, $\delta = 2\fdg5$, no cuts in
optical brightness;
2 -- optimization in $\Cr$ with $\delta = 2\fdg5$ and no cuts in optical brightness;
3 -- optimization in both $\Cr$ and $\Co$ with $\delta = 2\fdg5$;
4 -- optimization in $\Cr$, $\Co$, and $\delta$.
\label{fig1}}
%\end{figure}
\bigskip

We just demonstrated that
using the most straightforward assumptions, blindly, one can hardly find
the correlation signal. Regarding more specific cuts, like in TT01,
one meets a problem of interpretation of the signal whether it is real or
is just a consequence of cuts optimization (see also  \citealt*{efs03,efs04}).
The claim, that a given cut was
motivated independently rather than was optimized, is not convincing unless
the motivation has been done {\it a priori}.

Now let us demonstrate how the multiple cuts optimization can actually mimic
a significant signal. In this demonstration we use $10^4$ random UHECRs samples
prepared as described in Sect.~\ref{sec:sample} and the BL Lac catalog with cuts, optimized
for each random sample. Fig.~\ref{fig1} shows the fraction of random samples $\eta$ which
demonstrated a ``significance of correlation'' higher than $p$, after cuts
optimization.
If we fix all the cuts (curve 1), then there is an approximate agreement between
$\eta$ and $p$.   If we optimize one cut, $\Cr$, then we obtain $\eta$ a
few times greater than $p$ (actually, the ratio $\eta/p$ can be interpreted
as the penalty factor discussed above). With two cuts optimization,
adding a scan over visual magnitude, the ratio $\eta/p$
reaches almost two orders of magnitude and one out of 5 samples demonstrates $p<0.01$.
If we add an optimization for the correlation angle $\delta$, then every third
random sample demonstrates a ``significance'' of $10^{-2}$,
every tenth gives $p<10^{-3}$, and
one out of thousand gives $p=10^{-6}$!

\section{Summary}

 We presented a method of a blind search for a hypothetic effect where
various trials   with different sub-samples or model parameters
do not affect the stated significance level.
%Such trials are unavoidable if one does not know exactly
%which fraction of data should display the effect most prominently.
We believe that a tradition to use this method, when possible, would dramatically
reduce the number of unconfirmed claims of marginally significant effects.
 The method is especially  useful when:
(i) there is a clear null hypothesis and a way to prepare random samples
representing it; (ii) there exists a convenient measure of the statistical
significance of the effect; (iii) the effect is uncertain in some respects,
otherwise a test with the blind {\it a priori} formulation (i.e. it is
{\it a priori} clear which data should be used and how the effect should look)
is sufficient.
 Such problems as searches for cross-correlation between two classes of
astrophysical objects usually satisfy all three conditions.
 We would like to emphasize that the proposed method is, in principle,  applicable in
any field of science.

 In this work, we performed a demonstration for only one size of the
array of random samples. To find the significance
level of the effect, one should make several trials with different array size
starting from a larger one, then reducing its size until the real sample
is found. The examiner should not disclose the real sample after unsuccessful
trial.

An effect detected with this method is credible because it ensures
a researcher against unintentional  overestimation of the significance.
The only possible source of errors that can mimic a positive result is a wrong
null hypothesis distinguishing random samples from the real sample.
In the case considered in this paper, this could be for example
a wrong exposure function of the UHECR detector. Otherwise, a positive result
would have an explicit meaning: the chance that the effect does not exist is
the inverse of the size of array of samples at the successful search.

As an application of the proposed method, we analyzed a possible
UHECR--BL Lac correlation. We found no significant correlation, but
cannot claim, of course, that correlation does not exist.

\acknowledgments
%\section*{Acknowledgments}

We are grateful to P. Tinyakov and I. Tkachev for useful discussions
and to the anonymous referee for numerous useful suggestions.
The work is supported by the RFBR grant 04-02-16987, the
Academy of Finland grants 80750 and 102181, the
Jenny and Antti Wihuri Foundation, the
Vilho, Yrj\"o and Kalle V\"ais\"al\"a Foundation, and the NORDITA
Nordic project in High Energy Astrophysics.

\end{document}